# Understanding the effect of drying time in process-structure-performance relationships for PM6-Y6 organic solar cells

Marc Steinberger[1], Maxime Siber[2,3*], Hans-Joachim Egelhaaf[1,2], Mingjian Wu[4], Irene Kraus[4], Johannes Will[4], Xianqiang Xie[5], Laju Bu[5], Jonas Graetz[6], Tobias Unruh[6,7], Larry Lüer[1], Erdmann Spiecker[4,7], Andreas Distler[1*], Jens Harting[2,3,8*], Christoph J. Brabec[1,2,9,10*], Olivier J.J. Ronsin[2]

[1]Institute of Materials for Electronics and Energy Technology (i-MEET), Friedrich-Alexander-Universität Erlangen-Nürnberg, Martensstraße 7, 91058 Erlangen, Germany

[2]Helmholtz Institute Erlangen-Nürnberg for Renewable Energy (HIERN), Forschungszentrum Jülich, Fürther Straße 248, 90429 Nürnberg, and Immerwahrstraße 2, 91058 Erlangen, Germany

[3]Department of Chemical and Biological Engineering, Friedrich-Alexander-Universität Erlangen-Nürnberg, Fürther Straße 248, 90429 Nürnberg, Germany

[4]Institute of Micro- and Nanostructure Research (IMN) & Center for Nanoanalysis and Electron Microscopy (CENEM), Department of Materials Science, Friedrich-Alexander-Universität Erlangen-Nürnberg, Cauerstraße 3, 91058, Erlangen, Germany

[5]School of Chemistry, Xi'an Jiaotong University, Xi'an 710049, China

[6]Institute for Crystallography and Structural Physics (ICSP), Friedrich-Alexander-Universität Erlangen-Nürnberg, Staudtstraße 3, 91058 Erlangen, Germany

[7]Interdisciplinary Center for Nanostructured Films (IZNF) and Center for Nanoanalysis and Electron Microscopy (CENEM), Friedrich-Alexander-Universität Erlangen-Nürnberg, Cauerstraße 3, 91058, Erlangen, Germany

[8]Department of Physics, Friedrich-Alexander-Universität Erlangen-Nürnberg, Fürther Straße 248, 90429 Nürnberg, Germany

[9]Institute of Energy Materials and Devices – Photovoltaics, Forschungszentrum Jülich, Immerwahrstraße 2, 91058 Erlangen, Germany

[10]Friedrich-Alexander-Universität Profile Center Solar (FAU Solar), Friedrich-Alexander-Universität Erlangen-Nürnberg, Martensstraße 7, 91058 Erlangen, Germany

*Corresponding Authors: m.siber@fz-juelich.de, andreas.distler@fau.de, j.harting@fz-juelich.de, christoph.brabec@fau.de

First authorship: Marc Steinberger and Maxime Siber

Abstract:

Making solution-cast organic solar cells industrially available generally comes at the cost of significant performance losses compared to device prototypes manufactured under laboratory conditions. Adjusting solvent evaporation kinetics is postulated to recover efficiency. Yet, a comprehensive characterization of their effect, independently of other property-defining parameters, is lacking. Thus, the present objective is to isolate the influence of the solvent drying rate on solution-deposited organic active layer nanomorphologies and performances. To this end, a specially designed gas quenching technique is employed to fabricate PM6:Y6 donor-acceptor films under systematic variations of evaporation conditions. Using an extensive investigation protocol that combines insights from numerical simulations and experimental measurements, process-structure-performance relationships are unraveled. It is found that higher drying rates imply finer and more dispersed nanomorphologies with increased fractions of amorphous material. This enhances electric charge generation, thereby improving short-circuit current density and overall cell performance. The open-circuit voltage is also boosted under accelerated evaporation due to changes in the aggregation mode of the Y6 small molecule that induce higher effective bandgaps. The results demonstrate that the developed gas-quenching technique is a valuable tool for optimizing performance of upscaled organic photovoltaics, as it is readily compatible with high-throughput equipment, such as roll-to-roll coating machines.

1. Introduction

Organic semiconductors are pivotal materials for various (opto-)electronic devices, including organic light-emitting diodes (OLEDs)[1–5], organic field-effect transistors (OFETs)[6–11], supercapacitors[12–15], and organic solar cells (OSCs)[16–21]. The latter, an emerging solar cell technology, has recently achieved power conversion efficiencies (PCEs) exceeding 20% on laboratory cells[22,23,23,24] and 14.5% on large areas[25,26]. These materials stand out for their ability to be solution-processed into films using roll-to-roll (R2R) techniques on lightweight, flexible polymer substrates, making them both cost-effective and versatile[18,27,28]. Moreover, their production is energy-efficient, avoids rare or toxic materials, and can utilize green solvents, which is beneficial to both reduce environmental impact and improve scalability for industrial applications.

The morphology of the dried semiconductor films is crucial for the device performance and is thus a significant research focus in organic electronics[29–31]. Achieving the optimal morphology is especially critical in organic photovoltaics (OPVs), as the final structure is a kinetically quenched state (rather than a thermodynamic equilibrium of the donor and acceptor materials) to facilitate charge generation and separation in the active layer[32–40]. For instance, multiple studies on the Poly[(2,6-(4,8-bis(5-(2-ethylhexyl)-4-fluorothiophen-2-yl)-benzo[1,2-b:4,5-b']dithiophene))-alt-(5,5-(1',3'-di-2-thienyl-5',7'-bis(2-ethylhexyl)benzo[1',2'-c:4',5'-c']dithiophene-4,8-dione))]:2,2'-((2Z,2'Z)-((12,13-bis(2-ethylhexyl)-3,9-diundecyl-12,13-dihydro-[1,2,5]thiadiazolo[3,4-e]thieno[2'',3'':4',5']thieno[2',3':4,5]pyrrolo[3,2-g]thieno[2',3':4,5]thieno[3,2-b]indole-2,10-diyl)bis(methanylylidene))bis(5,6-difluoro-3-oxo-2,3-dihydro-1H-indene-2,1-

diylidene))dimalononitrile (PM6:Y6) OPV model system highlight the importance of tailoring the active layer morphology to obtain satisfactory PCEs [41–53].

Morphological aspects are mostly investigated on devices manufactured in research laboratory environments, neglecting production constraints related to large-scale commercialization. On the route towards industrial upscaling, key challenges include the use of eco-friendly solvents, the fabrication of thick active layers, and the solution-deposition under ambient atmospheric conditions[54]. These factors may negatively affect morphology, necessitating precise control strategies such as additives[47,50,55], solvent mixtures[56], and high-temperature casting[43,55].

Zhao et al.[43] demonstrated that the processing temperature is a decisive parameter for morphology tuning. By replacing halogenated, lower-boiling-point solvents like chlorobenzene (CB) with higher-boiling-point ones such as o-xylene (OX) or 1,2,4-trimethylbenzene (TMB), they achieved comparable OPV efficiencies for PM6:Y6 films cast using hot-slot die coating. Relying on in-situ characterizations of morphology formation kinetics during film drying, they suggested that adjusting temperature, so that the evaporation rates of the different solvents match, permits to realize similar active layer morphologies and, hence, analogous device performances.

However, most of the critical thermodynamic parameters, which influence morphology formation (e.g., the driving forces for crystallization, the compatibility of the organic molecules in the amorphous state, the solubility of the donor and acceptor materials within the solvents, and, by that, the propensity of the solutes to preaggregate in solution), bear distinct non-trivial temperature-dependencies. So do the relevant kinetic parameters (that is, molecular self- and interdiffusion properties, the viscosities, the solvent evaporation rate, etc.). To further improve the morphology control upon upscaling, the aim of the present study is therefore to examine the Process-Structure-Performance (PSP) relationship between the solvent evaporation rate and the optoelectronic properties of the final device at constant temperature, i.e., independently of other variables.

For this purpose, gas-assisted drying, also referred to as gas quenching, is used in this work on wet organic active layer mixtures. Previously, only a few research groups in the OPV field have addressed the topic of gas-assisted drying[57–60]. Here, a dedicated experimental setup is constructed similar to the works of Spooner et al.[57], who used air-knife-assisted spray coating to fabricate PM6:DTY6 films from O-xylene (OX), and Cheng et al.[61], who improved the efficiency of CB-cast organic solar modules with a nitrogen-blowing approach. A key focus of this research is the upscaling of OSC production techniques for industrialization. The instrument is therefore designed so that it can be transferred without major modifications to factory equipment such as roll-to-roll printing machines. The device consists of an air knife mounted on a doctor blade, as detailed in the Methods section of the Supplementary Information (SI, part 5). The pressure and the gas flow can be precisely adjusted along with the distance of the nozzle to the drying film. Drying rate variations can then be undertaken by modulating the gas flow, which does not require a change in the processing temperature.

To obtain deeper insights into the PSP relationship related to the evaporation kinetics, this study explores a broad range of drying rates of PM6:Y6 films cast from three distinct solvents with varying solubility properties at their commonly employed processing temperatures. Chloroform (CF) is chosen as the standard low-boiling-point (61.2°C) halogenated solvent used for well-performing PM6:Y6 bulk heterojunctions fabricated in laboratories. In contrast, the less toxic and more environmentally friendly OX and 1-Methylnaphthalene (MN) are selected because of their higher boiling points (144.4°C and 245°C) and their relevance for industrial applications like large-scale high-throughput manufacturing[25,54,62] or inkjet printing[63].

It is first analyzed how the cell efficiency changes when the active layer is deposited under controlled air flow variations at fixed temperatures. Drift-Diffusion calculations are conducted to relate key performance indicators (i.e., short-circuit-current density, open-circuit voltage and fill factor) with optoelectronic properties. Then, it is sought to delve into the underlying morphological modifications provoked by the drying modulations. Thereby, a coupled theoretical-experimental investigation protocol is followed to benefit from both established knowledge in the field of material science and insights from experimental observations.

A recently developed Phase-Field framework is first employed to simulate evaporation-induced morphology formation in a simplified PM6:Y6:OX model representation[64–68]. Atomic Force Microscopy (AFM) measurements are subsequently carried out to assess surface roughness and phase locations in dry PM6:Y6 films. Film-depth-dependent Light Absorption Spectroscopy (FLAS) is used to profile the vertical distribution of the species in the photoactive layer. Additionally, Transmission Electron Microscopy (TEM) experiments are performed to gain further information about donor- and acceptor-rich regions, domain sizes, and privileged molecular stacking orientations. Ultraviolet-Visible (UV-Vis) spectroscopy is also employed to identify the aggregation behavior of the blended materials. All morphological features are compared against the device performance data (i.e., power conversion efficiencies, short-circuit current density, open-circuit voltage, and fill factor) to pinpoint relationships between processing, structure, and performance.

This study distinguishes itself from previous works by explicitly isolating and precisely controlling the drying rate through gas-quenching, thus allowing a direct investigation of performance changes induced by evaporation kinetics, independently from other temperature-driven effects. This helps to understand and improve morphology adjustment for OPV upscaling and permits to develop a convenient and cheap roll-to-roll-compatible morphology control system.

## 2. Results

### 2.1 Solar cell performance:

To examine the effect of the drying rate of wet active layer films on OPV device properties, solar cells are produced with solution-deposited active layers using various solvents (i.e., CF, OX, and MN) at several gas quenching rates (see Figure 1). The corresponding drying rates are estimated quantitatively using the well-established Hertz-Knudsen evaporation theory[69,70] (see SI, part 2, for more details). Thereby, the drying rates depend on the vapor pressures of the solvents, whose temperature-dependency is calculated using Antoine's equation. This simple approach provides reliable drying rate estimates, as confirmed by the comparison with White Light Reflectometry (WLR) measurements on pure solvent films (see SI, part 2). Regarding the effect of the present gas quenching technique, it is found that the drying rate varies exponentially with the applied gas flow rate (see SI, part 2).

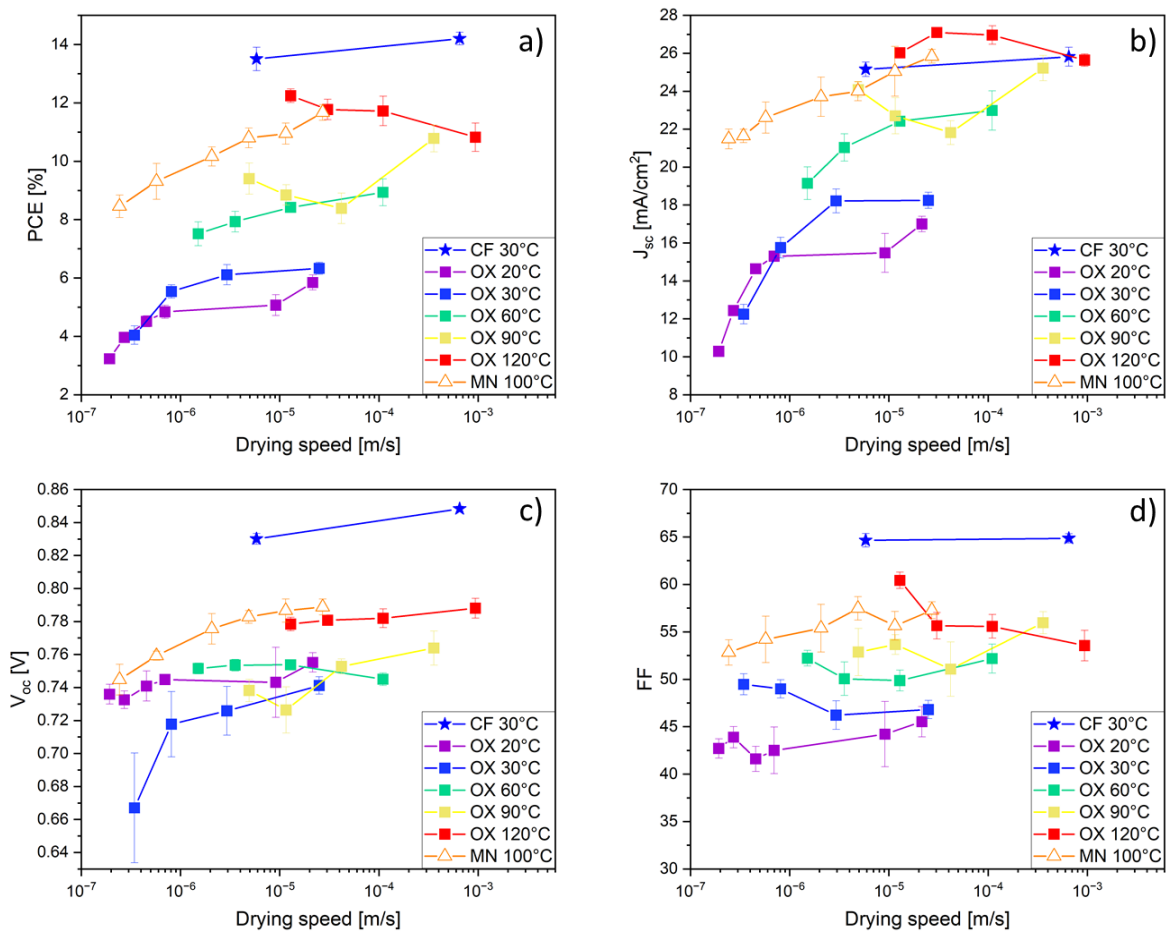

*Figure 1: Key performance indicators of PM6:Y6 organic solar cells (100 nm active layer thickness), prepared with chloroform (CF – stars), o-xylene (OX – squares) and methylnaphthalene (MN – triangles) as solvents, at different temperatures (20, 30, 60, 90, 100 and 120°C, as displayed in the legend) and with different drying speeds (calculated from the applied gas flow with the Hertz-Knudsen theory of evaporation, as detailed in the SI, part 2). a) power conversion efficiency (PCE), b) short-circuit current density ($J_{sc}$), c) open-circuit voltage ($V_{oc}$), d) fill factor (FF).*

Figure 1 shows the measured evolution of standardly investigated solar cell performance indicators (i.e., power conversion efficiency, short-circuit current density, open-circuit voltage, and fill factor) as a function of the speed of the liquid-vapor interface of the film during solvent evaporation (calculated with the Hertz-Knudsen theory and herein also referred to as "drying rate"). As mentioned, the characteristics of devices cast from CF, OX, and MN are compared. In the OX case, a processing temperature screening is performed to validate previous results by Zhao et al.[43] on efficiency recovery with high-temperature deposition. Note that the solar cells produced for this study are not optimized to yield champion performances, since the focus is to gain a detailed understanding of the effect of the gas quenching technique. More specifically, it can be pointed out that no thermal annealing post-treatment is performed to prevent an alteration of the properties of the as-cast active layers, which are sought to be compared here.

The study clearly reveals that the choice of solvent, processing temperature and gas quenching rate significantly influences power conversion efficiencies (PCEs), ranging from 3% for OX at RT to 14% for CF at 30°C (Figure 1a). For standard coating temperatures, CF (30°C) exhibits the highest PCEs compared to the higher boiling solvents MN (100°C) and OX (60°C). As demonstrated for OX, increasing the temperature can lead to a substantial performance increase thus confirming the findings of Zhao et al.[43] At any fixed temperature, and by incrementing the drying speed only, the PCE also increases in most cases, except for OX at 120°C, for which it decreases. For OX at 90°C, a non-monotonous PCE trend is measured. Nevertheless, the efficiency is still globally enhanced between the natural drying and fastest evaporation conditions. Figure 1a demonstrates that the effect of a higher processing temperature is not synonymous for that of a higher drying rate, as comparable drying rates enforced under different temperature conditions do not result in the same device efficiency.

The main contribution to the PCE increase with drying speed stems from the short-circuit current density ($J_{sc}$). For OX, the increase of $J_{sc}$ with drying rate is more significant for lower temperatures than for medium and higher ones. At 120°C, gas quenching past an optimal drying speed range (around $3 \times 10^{-5}$ m/s) even slightly reduces the $J_{sc}$ with accelerated evaporation. The initial increase of $J_{sc}$ at low gas flow rates is approximately linear in the logarithmic presentation of Figure 1b. At higher gas flow rates, the $J_{sc}$ values tend to reach a plateau. The level of the $J_{sc}$ plateau is temperature-dependent, around 15 mA/cm² $J_{sc}$ for 20°C OX samples and 23 mA/cm² for 60°C OX samples, while 120°C OX samples have a maximum $J_{sc}$ of 27 mA/cm². Morphological causes responsible for the observed $J_{sc}$ plateau are further discussed in subsequent sections. It can additionally be remarked that similar $J_{sc}$ values (~ 25 mA/cm²) are achieved at a drying speed of around $10^{-5}$ m/s for CF, MN, and OX samples, cast respectively at 30°C, 100°C, and 120°C. This highlights that combining the gas quenching technique with high-temperature processing allows recovering (and even outperforming) the short-circuit current density performance obtained from low-boiling-point, halogenated solvents, such as CF.

The $V_{oc}$ values shown in Figure 1c increase as well with faster drying speeds and are also temperature- and solvent-dependent. Among all solvents, CF shows the highest $V_{oc}$ (0.83 V), which increases even further upon gas quenching (0.85 V). MN displays $V_{oc}$ values which gradually increase with drying speed (between 0.74 V and 0.79 V). Finally, for OX, the $V_{oc}$ is also generally increasing with the drying rate, although the trend is less clear, as the curves have different slopes and are not always monotonous. Interestingly, the temperature dependence is here different than that observed for the PCE and the $J_{sc}$. Indeed, regardless of the temperature, the $V_{oc}$ is comprised between 0.73 V and 0.76 V, except for two outlying cases. The first exceptions are the OX samples deposited at 30°C under slow evaporation (that is, below a drying speed of $10^{-6}$ m/s), which display lower $V_{oc}$ (however, with a relatively high variance, so that the upper bound of the measurements approaches the values obtained at other processing temperatures). Nonetheless, when higher gas quenching rates are applied, the $V_{oc}$ lies within the aforementioned range. The second exceptions are all the devices produced at 120°C, which exhibit enhanced $V_{oc}$ values around 0.78 V to 0.79 V. Notably, the $V_{oc}$ of these solar cells is comparable to that of devices fabricated with MN at similar drying rates.

Finally, the FF does not exhibit uniform variations with drying speed: The FF stagnates at 65% for CF, which is the highest recorded value among all solvents. In comparison, an FF increase (from 52% to 57%) is visible for MN solar cells with accelerated evaporation. The FF measured for OX-cast samples decreases slightly at intermediate drying rates but rebounds at the highest ones. The only exceptions are the active layers deposited at 120°C, for which the FF steadily decreases. Note that, in contrast to higher drying rates, higher temperatures consistently lead to an FF augmentation for OX samples. This points out again that the effect of an increased processing temperature cannot be reduced to that of an increased drying rate.

The results show that accelerated drying is capable of improving solar cell performance for all three solvents, indicating that it indeed has a major effect on the active layer morphology and, thus, on the electrical properties. However, while Zhao et al.[43] reported the same efficiencies for solar cells cast from chlorobenzene (CB) and 1,2,4-trimethylbenzene (TMB) at high temperature, this could not be reached here for CF against OX (or MN) by solely matching the drying speeds, as CF still presents higher metrics. A first important conclusion of this study is that drying rate adjustments give rise to different changes as compared to temperature control, even though there is a direct relation between any solvent's natural drying speed and the temperature. The scope of this paper is the elucidation of the process-structure-performance relationships for the drying rate distinct from other process variables. Therefore, the investigations of the upcoming sections concentrate on active layer property changes implied by drying rate variations at fixed processing temperatures.

## 2.2 Optoelectronic property changes with accelerated drying

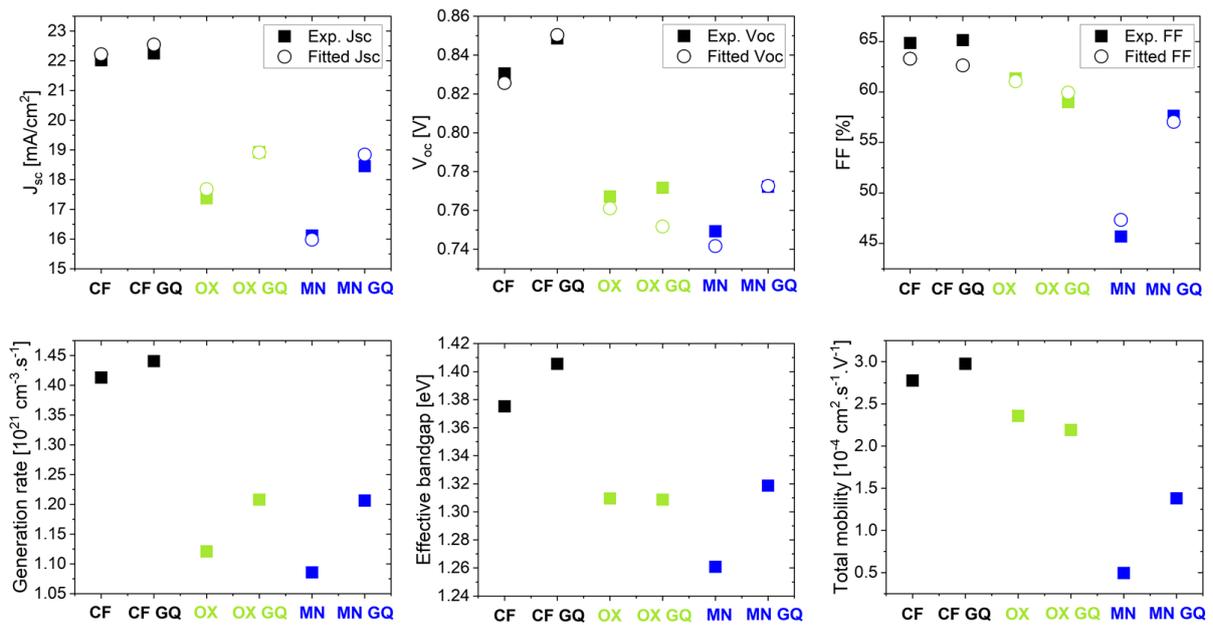

Figure 2: Results of Drift-Diffusion simulation fits carried out for IV curves stemming from PM6:Y6 OSCs (100 nm active layer thickness) produced from CF (at 30°C), OX (at 65°C), and MN (at 100°C) under natural drying and gas quenching (55 l/min air flow). The upper row presents the comparison of the measured and the simulated IV curves, as represented by a) the short-circuit current density, b) the open-circuit voltage, and c) the fill factor. The lower row shows the fitting parameters that exhibit the strongest correlations with the trends depicted above, namely d) the free charge carrier generation rate, e) the effective active layer bandgap, and f) the total effective free charge carrier mobility (i.e., the sum of both hole and electron mobilities).

In order to gain a better comprehension of the connection between the performance indicators and the optoelectronic properties, Drift-Diffusion simulations are now performed and fitted to JV curves of PM6:Y6 solar cells fabricated with CF at 30°C, OX at 65°C, and MN at 100°C. Figure 2 presents the most notable variations of the optoelectronic parameters with drying rate and solvent type, as obtained from the fitting procedure. The fitted JV curves along with more details about the employed methodology can be found in the SI (SI, part 4). Globally, the performance indicators from the Drift-Diffusion fits match with the experimental ones. A slight difference in $V_{oc}$ (~2% relative error) can be noticed for the gas-quenched OX sample. Nevertheless, a satisfactory agreement is obtained between all experimental and simulated JV curves over the whole investigated voltage range, as shown in the SI (SI, part 4).

As expected, the short-circuit current density (Figure 2a) is mainly correlated with the charge generation rate (Figure 2d). These variations of the generation rate cannot be justified only by changes of the absorption properties with the processing conditions (see the absorption spectra of OSCs produced under natural drying and gas quenching conditions in Figure 12). This suggests that the increase of $J_{sc}$ under gas quenching is mainly due to an enhanced exciton dissociation efficiency.

The $V_{oc}$ increase upon accelerated drying for CF and MN (Figure 2b) seems to be fully related to an increase in the effective bandgap of the active layer (Figure 2e). This is in line with the blue shift observed in the absorption spectra under gas quenching, as detailed more extensively in the section dedicated to the spectral analysis (Figure 12). For OX at 65 °C, for which no strong gas flow rate dependence of the $V_{oc}$ is measured, the effective band gap remains constant.

In general, the FF of a device is affected by hole and electron mobilities, as well as by Langevin and Shockley-Read-Hall (SRH) charge carrier recombination rates[71–73]. A direct correlation of these parameters with the FF is not observed. Nevertheless, it can be pointed out that gas quenching tends to slightly increase the Langevin recombination rate as well as the SRH trap density for all devices independently of the processing solvent (see SI-Figure 14). However, the optoelectronic quantity found to correlate best with the FF variations (Figure 2c) is the sum of hole and electron mobilities (labelled here as "total mobility" in Figure 2f). This means that this overall mobility is determinant for the FF in the present case, which agrees with previous studies[74]. The total mobility (and, thus, the FF) does not vary much upon gas quenching, except for MN samples, for which it leads to notable improvements.

To further understand how solvent choice and drying speed influence the electrical properties of the active layers, additional analyses are carried out with transient photovoltage (TPV)[75], charge extraction (CE)[76,77], and charge extraction by linearly increasing voltage (photo-CELIV)[78–82] experiments (see SI-Figure 1). Accelerated drying generally results in slightly faster recombination, indicated by a reduction in charge carrier lifetime (τ), as observed in TPV and CE measurements. This is in line with the Drift-Diffusion fits for the Langevin recombination rate and the SRH trap density (see SI-Figure 14). Mobility changes inferred from photo-CELIV are minimal, suggesting a limited influence of the drying speed on charge transport properties. Collectively, these findings indicate that accelerated drying does not fundamentally alter the transport and recombination behavior in most cases, which explains why FF changes remain minor. Further details and supporting measurement data are provided in the SI (SI, parts 1 and 4).

In summary, accelerated drying primarily enhances the $J_{sc}$, slightly increases the $V_{oc}$, and leaves the FF largely unaffected. One-diode model analysis, supported by TPV, CE, and photo-CELIV measurements, links these effects to a pronounced increase in charge generation, a modest bandgap widening, and nearly unchanged charge carrier mobility and recombination, respectively.

### 2.3 Phase-Field simulations for evaporation-induced morphology formation visualization

To understand the effect of the drying speed on the photoactive film morphology formation, simulations of the solution-deposition process are carried out with an in-house computational framework. Extensive descriptions of its development and usage can be found in previous publications[64–68,83–85] and are therefore not included in this manuscript. Most importantly, the employed model relies on the Phase-Field approach, which permits to render multiphase systems and capture the interplay between various phase transformation mechanisms (e.g., phase separation, crystallization, evaporation) and mass transport (both diffusive and advective), all being of major relevance to understanding the phenomenology of the drying process. Another substantial advantage of Phase-Field modeling is that all the important timescales of the solution-processing are accessible with reasonable computational effort.

Although the primary aim is to study the influence of the drying rate qualitatively, particular care is taken to base on realistic input parameters to obtain an adequate representation of the PM6:Y6 bulk heterojunctions examined in this manuscript. A summary of the utilized thermodynamic and kinetic material parameters is provided in the supporting information (SI part 3) along with additional details concerning the simulation approach. One main simplification undertaken here, however, concerns the PM6, which is treated as a fully amorphous polymer, despite literature reports evidencing its aggregation behavior[40,43,86]. The reason is that representing its complex semi-crystalline nature properly requires detailed information about the melting thermodynamics and the crystallization kinetics of the PM6 in solution. The characterization of these fundamental aspects lies outside of the scope of this work and is thus reserved for further developments. While it cannot be left out that PM6 crystallization may be responsible for specific morphological features, it is not expected to alter significantly the general trends presented in this section.

The simulated system (Figure 3) consists of a wet film beneath its associated vapor phase. On the one hand, the film is composed of a polymer donor (PM6), a small molecule acceptor (Y6), and a solvent (OX) in a condensed phase, which can be either in the amorphous (liquid) or the ordered (crystallized) state. On the other hand, the vapor is composed of ambient air and solvent in gaseous form. An outflux of solvent calculated with the Hertz-Knudsen model (SI part 2) is applied at the top boundary of the simulation box. Since gas and condensed phases are in a quasi-equilibrium state, this leads to solvent evaporation at the liquid-vapor interface until the film is dry[65,66].

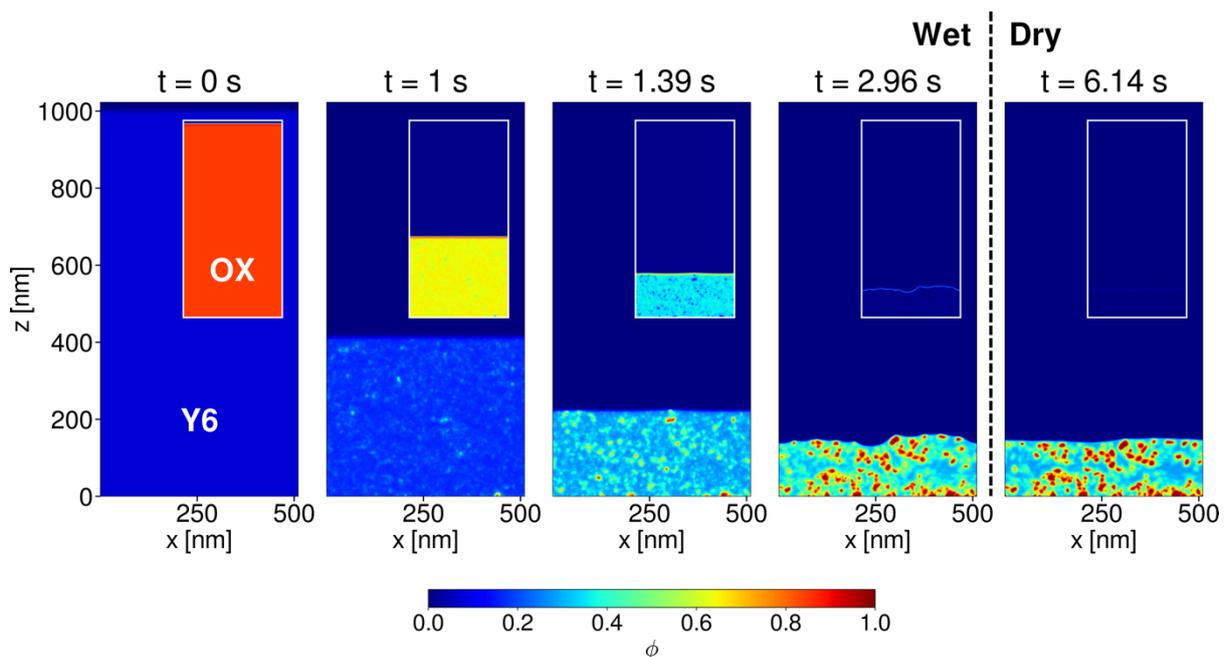

*Figure 3: Example of a two-dimensional Phase-Field simulation representing the solution-deposition process of a PM6:Y6 bulk heterojunction cast from o-Xylene (OX). The snapshots show the volume fraction fields ($\phi$) of Y6 (main figures) and OX (insets) at successive stages of the evaporation until the photoactive film is dry. The color code ranges from deep red at locations where the material is completely pure to deep blue where it is vanishing. The regions where the Y6 field progressively turns red are the sites where it crystallizes.*

Figure 3 depicts the concomitant progress of solvent evaporation and Y6 crystallization in a typical simulation case. At the beginning (t=0 s), both the donor and the acceptor are completely dissolved in the mixture. Due to solvent extraction, the PM6 and Y6 concentrations increase in the wet film (t=1 s to t=6,14 s). Once the critical concentration is reached for the small molecule (i.e., the limit Y6 fraction in the amorphous phase beyond which nucleation can be expected for the length- and timescales covered by the present system, see SI part 3 for more details along with Figure 4b), crystallization is in principle possible. First nucleation events take place beyond this threshold, depending on the evaporation kinetics (here at t=1,39 s). Further crystal nucleation and growth may occur (t=2,96 s) until the film is dry (t=6,14 s), provided the Y6 content in the amorphous phase remains both above the critical concentration for nucleation, and above the saturation concentration (i.e., the liquidus equilibrium composition of the phase diagram, see SI-Figure 8 and Figure 4b) for crystal growth.

The final morphology is the result of a competition between the drying and the crystallization (nucleation and growth) processes. The crystallization kinetics are strongly dependent on blend composition (see SI part 3). Globally, the phase transition requires a certain amount of time to proceed in the wet mixture before it is kinetically quenched in the dry film. This time is provided during solvent evaporation. In principle, slower drying thus allows more advanced crystallization. Nevertheless, below the critical concentration, crystallization cannot take place. Above the critical concentration, but still at relatively high solvent content, the mobilities of donor and acceptor molecules are high, but the thermodynamic driving force for crystallization is so low that the crystallization kinetics are slow, nucleation events are rare, and crystallization is growth-dominated. Conversely, at very low solvent content (towards a dry film), the thermodynamic driving force is strong so that the crystallization becomes nucleation-dominated. Nonetheless, the mobilities are low, so that its kinetics are also slow. In between, at intermediate solvent content, there is a sweet spot for fast crystallization, where both the driving force and the molecular mobilities are sufficiently high. The evaporation rate controls how fast the drying film progresses through the composition range from high to low solvent content. Therefore, it determines the supersaturation at which the first nuclei form, the overall balance between nucleation and growth, the global extent of the phase transition, and, consequently, the nanomorphology of the dry film.

Under slow evaporation, the supersaturation of Y6 in the mixed amorphous phase stays low over the course of the crystallization (see Figure 4b and SI-Figure 10c), and so does the thermodynamic driving force for crystallization. As a result, nucleation is weak and crystallization is growth-dominated. Few crystals appear and the Y6 material available in the amorphous phase is mainly consumed by crystal growth. Since crystallization occurs in a wet film with considerable solvent amount, the growth process is fast and has time to exhaust most of the amorphous Y6 before the end of drying (see Figure 4a), so that a high crystallinity (here around 95 % Y6 crystallinity, see Figure 4d) and large crystal sizes (e.g., up to a mean diameter of 38 nm in Figure 4c) are obtained. Upon increasing the drying speed, the Y6 content in the mixed amorphous phase from which nucleation proceeds is enhanced. This higher supersaturation (Figure 4b and SI-Figure 10c) yields a more nucleation-dominated crystallization process and thus a higher nucleus density. However, since crystallization also takes place at lower solvent content (Figure 4a and Figure 4b), molecular diffusion kinetics are accordingly slower. Therefore, the crystallization process has no time to complete before the end of the drying. The morphology is kinetically quenched. Thus, the dry film morphology retains more of the amorphous acceptor (i.e., only 30 % and 7 % Y6 crystallinity for the intermediate and fast evaporation rates, respectively, see Figure 4d), and the crystal sizes are smaller (about 13 nm and 6 nm mean diameter for medium and fast evaporation, respectively, see Figure 4c).

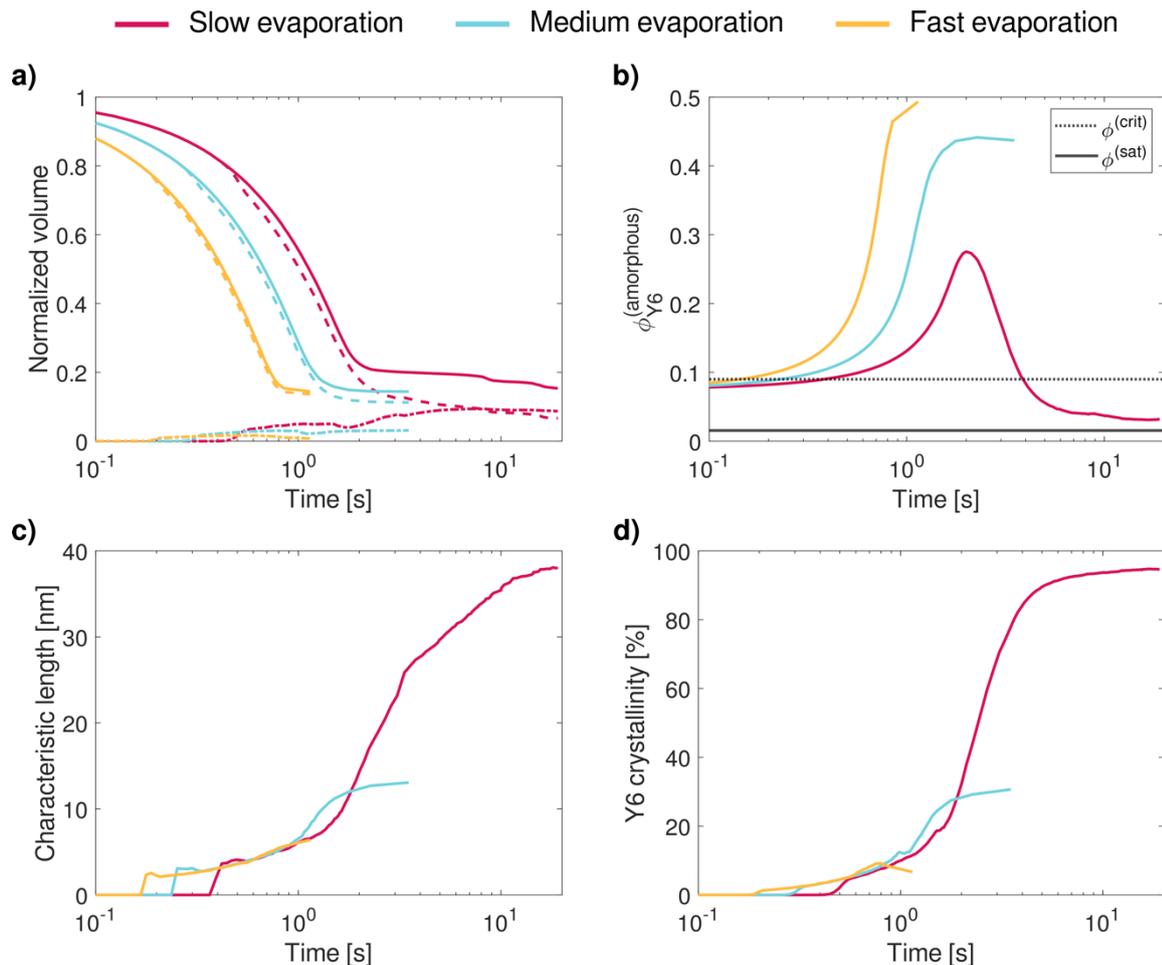

*Figure 4: Time-evolution of the properties of the drying PM6:Y6:OX films under slow, medium, and fast evaporation conditions. a) Normalized total film volume (plain lines), amorphous volume (dashed lines), and Y6 crystallized phase volume (dash-dotted lines). b) Y6 volume fraction in the amorphous phase of the film ("LaMer" curve). The volume fraction $\phi^{(crit)}$ indicates the critical concentration beyond which Y6 crystal nucleation can take place. The equilibrium, or saturation volume fraction $\phi^{(sat)}$ corresponds to the concentration above which the crystallization proceeds. c) Characteristic length of the Y6 crystals, defined as their average equivalent diameter. d) Y6 crystallinity.*

Figure 5a shows how the dry film morphology changes with varying drying rate (all other process conditions remaining constant), according to the physical mechanisms described above. A slow evaporation rate results in a dry film morphology composed of large, well-separated domains, with most of the Y6 material being crystalline. These coarse structures also imply a higher surface roughness of the final film. Moreover, it can be seen in the vertical composition profiles (Figure 5b), that a larger amount of Y6 acceptor is found in the vicinity of the liquid-vapor interface for low evaporation rates. This is because nucleation occurs when the volume of the wet film is still relatively high. Crystals appearing by homogeneous nucleation are initially randomly distributed in the thick liquid mixture. However, crystals reached by the moving liquid-vapor interface upon drying are advected towards the substrate due to the capillary forces it exerts on them. As a result, the crystals progressively gather at the top of the film[87]. This accumulation effect is responsible for the macroscopic phase segregation observed in the composition profiles at the film surface.

A larger amount of Y6 is also found at the bottom of the film, because heterogeneous nucleation at the substrate is energetically more favorable than homogeneous nucleation in the bulk of the drying film. A strong occurrence of heterogeneous nucleation is nevertheless not systematically expected in practical cases, the phenomenon relying strongly on the surface tension reduction induced by the employed substrate on the crystallizing component. For higher drying rates, along with lower crystallinity and smaller crystal sizes, the different domains in the final films (Figure 5a) exhibit a higher surface-to-volume ratio, a higher connectivity from the bottom to the top surface, and a lower phase purity (i.e., a higher mixing of both blend components). Additionally, the vertical structuring, which is strongly related to an early and advanced crystallization process, also progressively disappears with increasing drying rates.

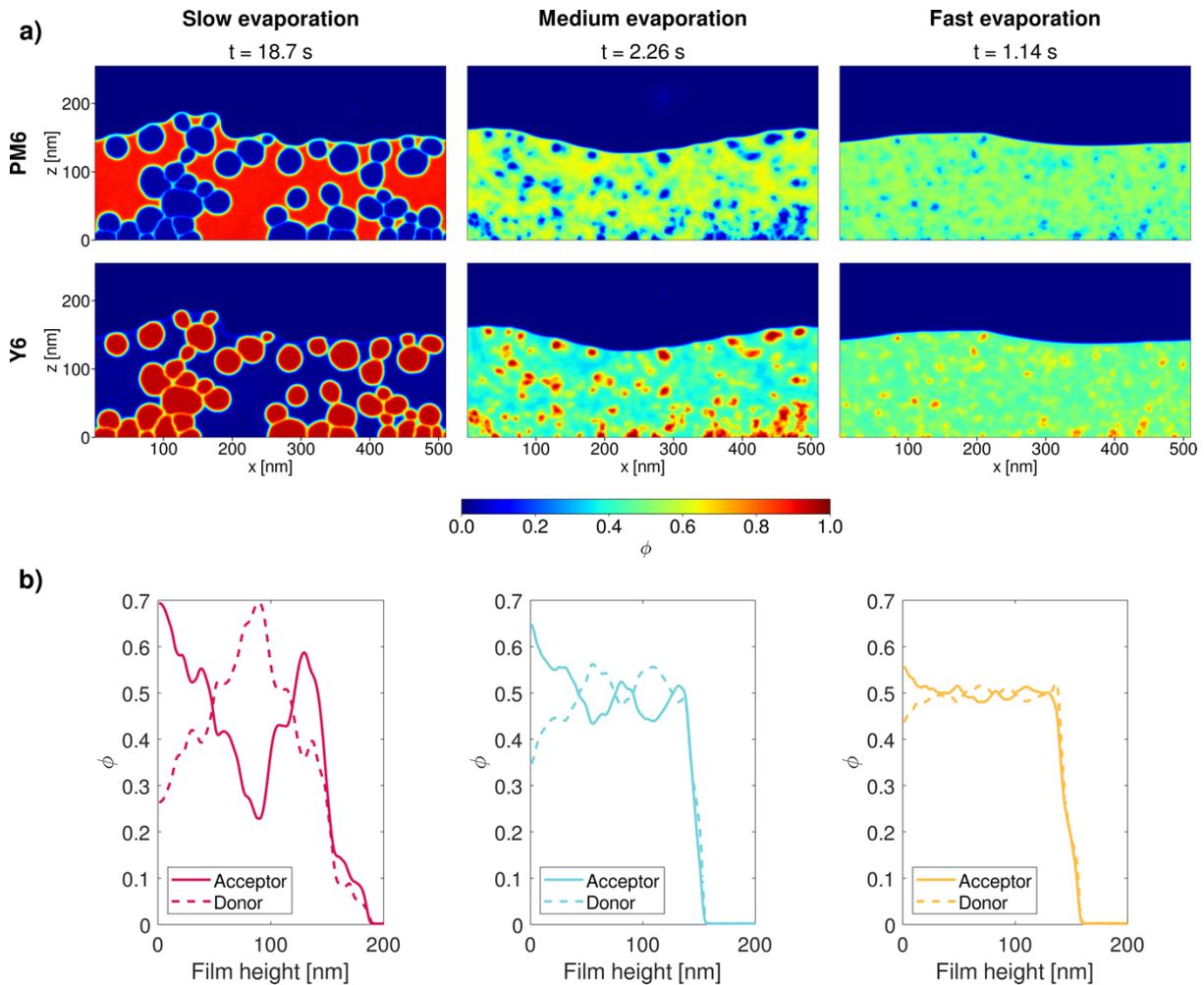

Figure 5: Dry film morphologies obtained in Phase-Field simulations of PM6:Y6 active layer deposition for slow (left), medium (center), and fast (right) evaporation rates. a) Volume fraction field (φ) showing the distribution of PM6 and Y6 material across the bulk heterojunction at the end of the drying. b) Corresponding vertical composition profiles (obtained by integration over the width of the simulation box for each height z).

All in all, the Phase-Field simulations demonstrate how controlling the evaporation rate can impact the morphology of the produced photoactive layer, even when the processing temperature is fixed. The main consequence of faster drying is a finer phase separation with higher intermixing, decreased crystallinity, and smaller crystal sizes (Figure *6*). The evaporation process can be accelerated to the point where all the solvent is removed before any aggregation process can occur. Therefore, there is a certain threshold beyond which increasing the evaporation speed further does not affect the morphology of the dry film anymore. The results thus suggest that there exists an optimal drying speed for which the mixed and separated, amorphous and crystalline phases form adequately-sized domains with satisfactory percolation properties. The aim is now to investigate this further with the help of dedicated experimental characterization techniques.

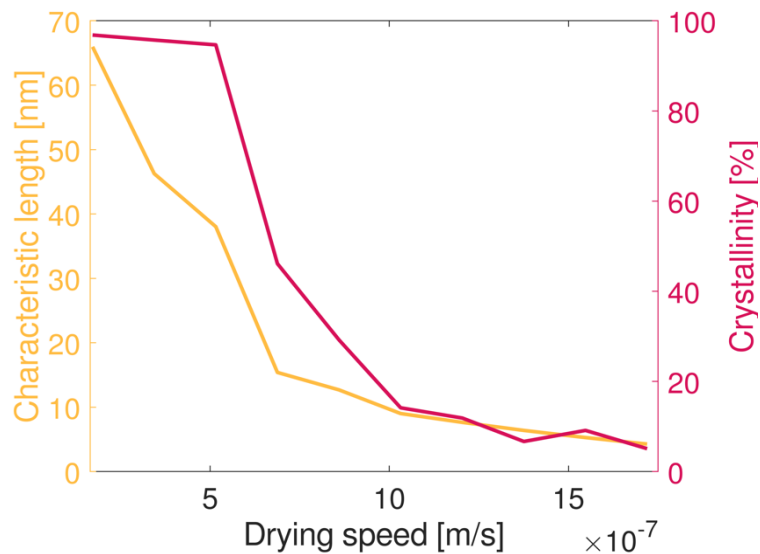

*Figure 6: Characteristic length of the Y6 crystals (defined as the average equivalent crystal diameter) and Y6 crystallinity in simulated dry PM6:Y6 films as a function of the drying speed.*

### 2.4 Experimental morphology analysis:

Atomic Force Microscopy (AFM) experiments are carried out on photoactive films cast from the three investigated solvents (at 30°C, 65°C, and 100°C, for CF, OX, and MN, respectively), both under natural and accelerated drying conditions. Figure 7 presents the corresponding phase mode images, which provide insight into the domain sizes and distributions of the different surface morphologies. The indicated root mean square (RMS) averages of the height profile deviations are calculated from the associated height maps included in the SI (SI-Figure 2).

The active layer produced from CF exhibits a relatively smooth surface (RMS = 0.8 nm), while those deposited with the slower-evaporating solvents display higher roughness (i.e., RMS = 5.2 nm for OX, and RMS = 2.9 nm for MN). Accelerating the evaporation process does not change the RMS for CF, slightly reduces it to 2.4 nm for MN, and halves it for OX (RMS = 2.6 nm). The effects on surface roughness induced by the transition from fast- to slow-evaporating solvents are mirrored by the phase images: The finest structures are obtained for CF, and the coarsest for OX and MN. Employing gas quenching does not significantly affect the domains visible in CF-cast films. In contrast, it causes a modest decrease in domain size for MN and a substantial one for OX. Both observations suggest that the nanostructure becomes finer when accelerating the drying of relatively slow-evaporating solvents, and that it does not notably evolve anymore at the fastest evaporation rates, confirming the expectations from the Phase-Field simulations.

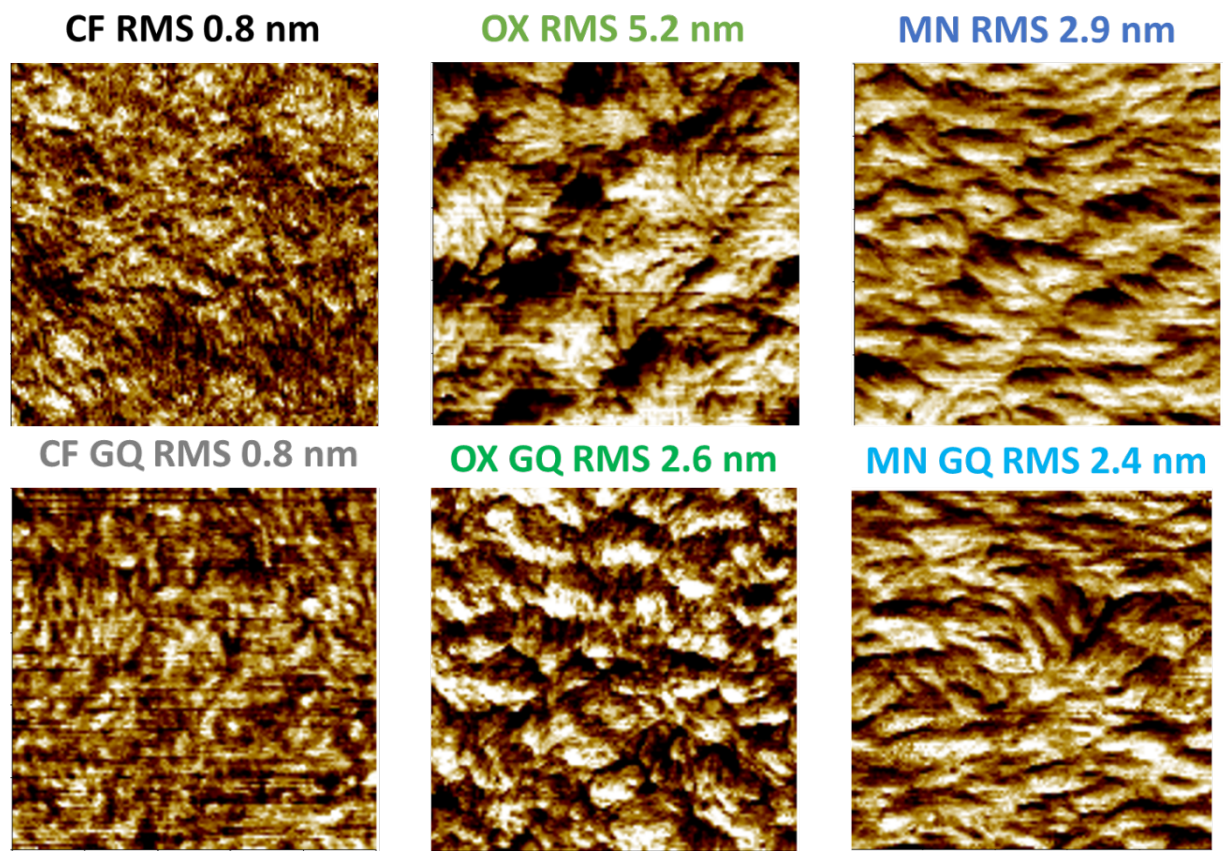

*Figure 7: 500nm x 500nm AFM phase images of different PM6:Y6 layers made from chloroform (black), o-xylene (green) and methylnaphthalene (blue) without (top row) and with accelerated drying (GQ – bottom row); RMS values given in the snapshots are calculated from AFM height information (see SI Figure 8).*

Information about the vertical composition profiles is provided by Film-depth-dependent Light Absorption Spectroscopy (FLAS)[88]. The CF-deposited samples without gas quenching show homogeneous vertical distributions of the donor and acceptor species, with only a slight excess of acceptor at the active layer/air (ALA) interface, indicating that the associated phases are well-mixed (Figure 8). In contrast, both the OX and the MN samples exhibit increased donor content immediately below the ALA interface, while in a region of around 20-70 nm below the surface, the situation is reversed. This suggests a macroscopic vertical segregation of the donor and acceptor materials, as obtained in the simulations at low evaporation rates (Figure 5b).

Gas quenching reduces this phase segregation, leading to more homogeneous films (Figure 8, bottom row), as predicted by the Phase-Field simulations (Figure 5). For MN, especially, the profile becomes comparable to that of non-gas-quenched CF-cast layers. It is noted that for all three solvents, accelerating the evaporation process results in an increase in the acceptor concentration at the ALA interface. For CF, this results in a somewhat more heterogeneous composition near the top interface (as well as at the bottom interface), which is not anticipated from the previously shown simulations. However, in the bulk of the active layer cast from CF, the most balanced composition ratio (i.e., the highest intermixing) is achieved for accelerated drying, as expected.

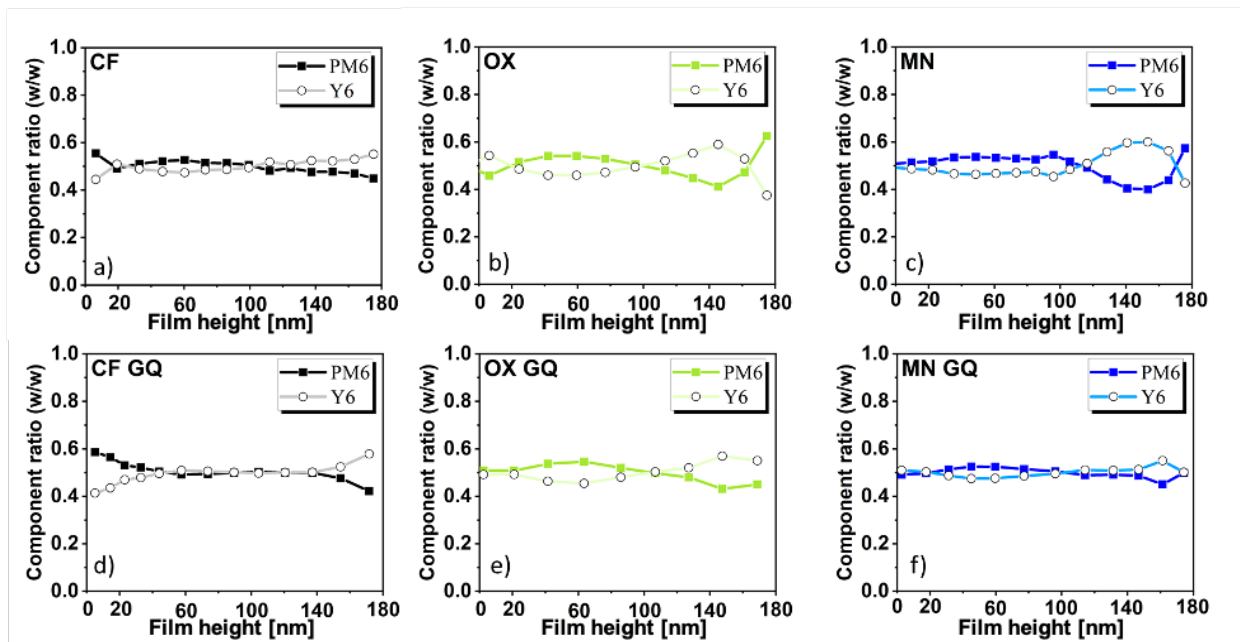

*Figure 8: Vertical phase separation profiles extracted from Film-Depth-Dependent Light Absorption Spectroscopy (FLAS) measurements on different PM6:Y6 layers made from chloroform, o-xylene and methylnaphthalene with and without accelerated drying.*

Transmission Electron Microscopy (TEM) is used to gain insights into the nanomorphology of the donor-acceptor heterojunction and the structure of the active layers. Electron energy-loss spectroscopy is first coupled to scanning TEM (STEM-EELS) to quantify the sample thickness and to map the carbon and sulfur elemental distributions. The contrast in the sulfur map reveals the distribution of PM6, as it contains more sulfur atoms per volume than Y6[89–91]. The results are summarized in Figure 9. In the sample processed from CF under natural drying (ND), the sulfur map shows a fine phase separation pattern with approximately 10 nm wide donor-enriched domains, separated by donor-depleted areas spanning over 30 to 50 nm. The gas-quenched (GQ) sample clearly displays a more homogeneous nanomorphology. In the ND OX case without GQ, the phase separation results in coarser donor-rich regions, approximately 30 nm in width, and donor-poor ones, ranging from 100 to 150 nm. With GQ, the nanostructure becomes much finer again, but not as smooth as for the CF samples. In the ND MN sample, in contrast, leaf-shaped PM6 regions with ~100 nm in length and ~40 nm in width are observed. Gas quenching the sample suppresses this feature, leading to a more homogeneous morphology that is comparable to the GQ OX one.

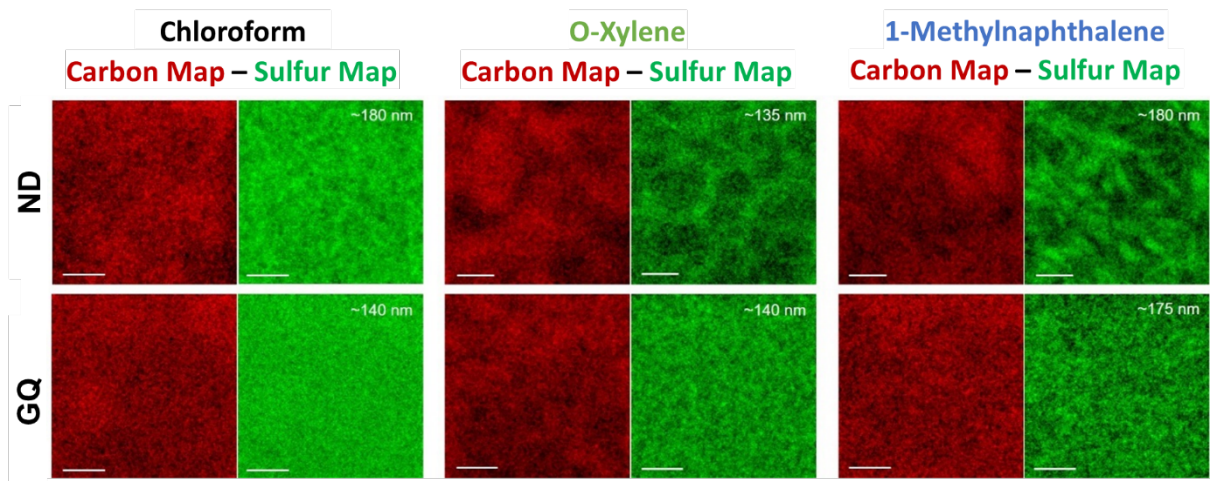

*Figure 9: Elemental maps (net signal) of carbon (red) and sulfur (green) of the series of samples (cast from CF, OX, MN at 30°C, 65°C and 100°C, respectively, using natural drying (ND) and gas quenching (GQ)) evaluated from STEM-EELS datasets. While the carbon maps predominantly reveal variations in thickness, the sulfur maps encode the distributions of PM6 (sulfur-rich) and Y6 (sulfur-poor). Numbers at the top right corner of each panel denote the sample thickness evaluated from the low-loss region of EELS data (cf. methods). Scale bars: 100 nm.*

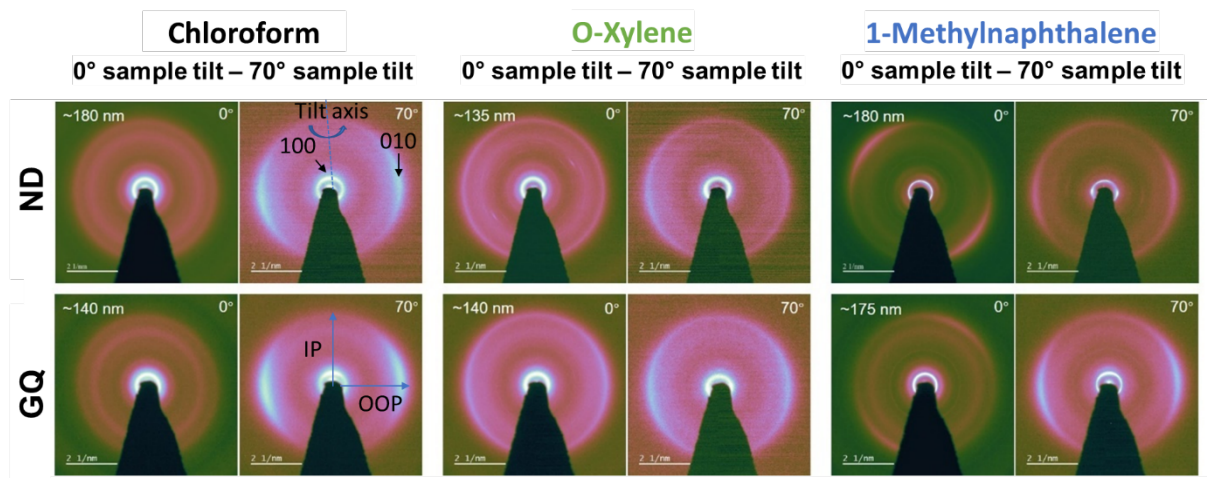

*Figure 10: Elastically-Filtered Selection Area Electron Diffraction (EF-SAED) of the sample series taken at 0° and 70° sample tilt (with respect to the incident electron beam), revealing both the in-plane (IP) and out-of-plane (OOP) structure character of the samples. The tilt axis is approximately aligned with the beam stopper, as marked by the dashed line on the top left panel. The lamella packing and π-π stacking diffraction peaks, i.e., the (100) peak at ~0.48 nm$^{-1}$, and the (010) peak at ~2.7 nm$^{-1}$, respectively, are marked. Note that the diffraction peaks could not be assigned to either one of the components, i.e., PM6 or Y6, as their characteristic packing spacings are expected to be very close and the ordered regions are small (thus, broad peaks overlap). It is therefore referred to the "edge-on" or "face-on" arrangement of the blends, globally, rather than that of the components, specifically.*

The character of in-plane and out-of-plane molecular ordering within the active layers is further unraveled using elastically filtered selection area electron diffraction (EF-SAED) at 0 and 70 degree sample tilt[89,90,92]. At 0°, the diffraction patterns carry only in-plane (IP) information, whereas the 70° tilted geometry allows the out-of-plane (OOP) information to be revealed at locations perpendicular to the tilt axis. The EF-SAED patterns of the sample series for qualitative examination are summarized in Figure 10. The extracted IP and OOP profiles for quantitative evaluation are presented in Figure 11 and SI-Figure 3.

Samples prepared from chloroform (CF) predominantly show a face-on orientation, meaning that the molecular planes are parallel to the substrate and π-π-stacking occurs vertically relative to the substrate. This is evident from the clear presence of characteristic 010 diffraction signals when viewed out-of-plane (OOP), and their absence in-plane (IP). Gas quenching does not alter this general orientation but slightly reduces the size of the ordered molecular regions, as indicated by broader diffraction peaks (see Figure 11b).

In contrast, samples from o-xylene (OX) display a mixed orientation of molecules, partially face-on and partially edge-on (π-π-stacking parallel to the substrate). Accelerated drying via gas quenching refines the nanostructure and slightly enhances face-on orientation, although edge-on orientation remains detectable.

Samples processed from methylnaphthalene (MN) without GQ exhibit the most pronounced edge-on molecular arrangement compared to the other solvent variations. Moreover, these samples reveal distinct long-range in-plane ordering. Gas quenching effectively reduces both the dominance and extent of the edge-on orientation, promoting a more balanced orientation similar to that observed in OX samples.

These findings show that solvent choice not only influences the domain sizes and morphology but also significantly impacts the orientation of the molecules within the films. Here, it is observed that CF favors face-on, OX promotes mixed face-on and edge-on and MN drives towards mixed with a slight tendency to edge-on and long-range IP orientation correlation. Gas quenching leads to broader diffraction peaks, which means smaller ordered regions — findings that align well with the Phase-Field simulations —, and favors face-on orientation.[93]

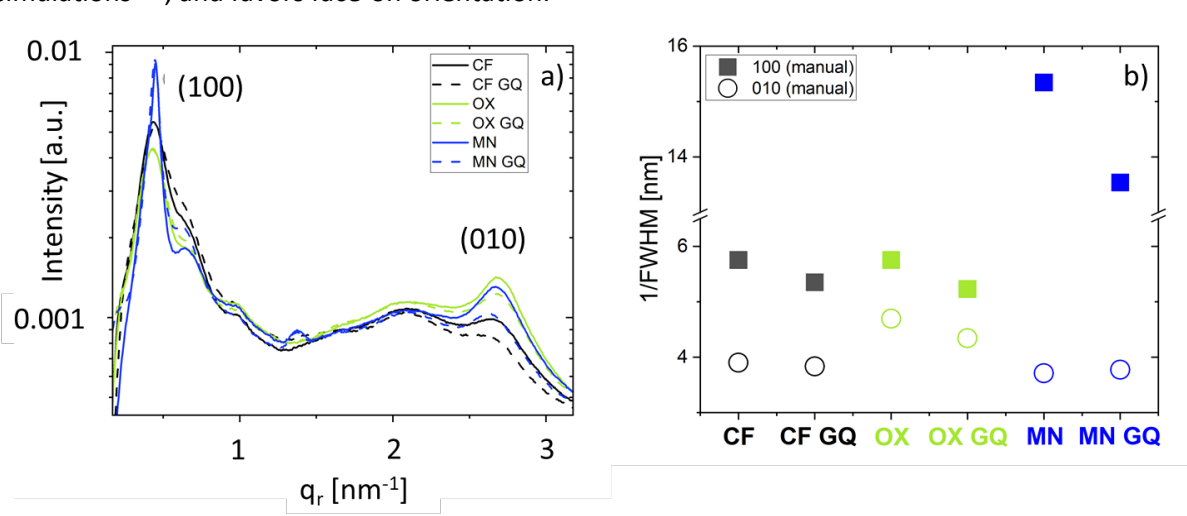

*Figure 11: Extracted diffraction profiles for quantitative evaluation of in-plane (IP) molecular ordering. a) IP diffraction profile from azimuth integrated EF-SAED at 0° tilt for the samples cast from CF (black), OX (green) and MN (blue), without (solid lines) and with (dotted lines) gas quenching for accelerated drying. b) FWHM evaluation of a).*

Figure 12a and Figure 12b presents the UV-Vis absorption spectra for active layers cast from CF, OX and MN, with and without accelerated drying. For all samples two absorption peaks are observed: the one at low energies (approximately between 1.4 and 1.6 eV) corresponds to the Y6 small molecule, while the high energy peak (between 2.0 and 2.2 eV) is associated with the PM6 polymer. The peak positions undergo distinct shifts depending on solvent and drying speed. In the layers cast from CF without GQ, the maximum of the absorption peak of Y6 (~1.55 eV) is blue-shifted by about 0.05 eV with respect to those observed in the layers cast from OX (~1.50 eV) and MN (~1.50 eV). Accelerated drying induces this peak to shift towards higher energies across all solvents. While the shift is almost negligible for CF, it is so pronounced for the MN GQ (1.53 eV)- and OX GQ (1.54 eV)-based films that the differences in peak positions between CF, OX, and MN are reduced to less than 0.03 eV after gas quenching. These shifts correspond to changes in the molecular aggregation state of Y6, which has been shown to adopt three distinct conformations by Kupgan et al.[94] and Kroh et al.[40], namely the amorphous phase, the so-called aggregate 1, and the aggregate 2. Aggregate 1 involves two overlapping molecule segments in the crystal lattice arrangement (i.e., either both core-core and terminal-terminal, or two core-terminal bonds) and aggregate 2 possesses only one intermolecular interaction (either terminal-terminal, or core-terminal), as displayed in Ref.[40]. Each conformation exhibits unique energy levels. Molecular dynamics (MD) calculations by Kupgan et al.[94] predict the smallest HOMO-LUMO offset for aggregate 1, followed by aggregate 2 and the non-interacting amorphous configuration, with optical gaps around 1.52 eV, 1.62 eV, and 1.73 eV, respectively.

Further deconvolutions of the spectra from Figure 12a and Figure 12b are undertaken with a modeling approach using the spectral agent fitting tool created by Luer and co-workers[95] and adapted according to the findings of Kroh et al.[40]. The results clearly demonstrate the formation of the three different Y6 phases in active layers cast from each solvent. Figure 12c and Figure 12f show the fitted spectra of MN and MN GQ, featuring the three green sub-bands on the left, which correspond to the three conformations of Y6 described above (see SI-Figure 4 for all other solvents). While the energies at which the respective peaks are located may slightly differ from the MD predictions, their order is recovered. The evaluation of the fitting parameters shown in Figure 12d and Figure 12e reveals that the peak height of aggregate 1 is lowest for CF compared to OX and MN, and is reduced with gas quenching for CF and MN, but not significantly for OX (Table 1). For all solvents, gas quenching additionally increases the fraction of the Y6 amorphous phase, in agreement with the results from the Phase-Field simulations.

*Table 1: Summary of UV–Vis absorption analysis parameters of PM6:Y6 films processed under different conditions. The reported positions of the base of the Y6 peak are recorded for the wavelength at which the absorption signal crosses a threshold value of 0.2 OD (see Figure 12b). The samples are ordered, so as to highlight the correlation between a redshift of the peak base position and an increase in aggregate 1 peak height (and, conversely, decreases in amorphous Y6 fraction and in $V_{oc}$).*

|  | CF GQ | CF | MN GQ | OX GQ | OX | MN |
|---|---|---|---|---|---|---|
| **Y6 peak base position [eV]** | 1.41 | 1.4 | 1.4 | 1.39 | 1.38 | 1.37 |
| **Aggregate 1 peak height** | 0.14 | 0.25 | 0.29 | 0.47 | 0.47 | 0.69 |
| **Amorphous fraction** | 0.88 | 0.83 | 0.76 | 0.75 | 0.71 | 0.68 |
| **$V_{oc}$ [V]** | 0.83 | 0.81 | 0.78 | 0.76 | 0.76 | 0.74 |

The overall larger fraction of Y6 arrangements with higher optical band gaps (i.e., amorphous phase and aggregate 2) explains the blue shift of the absorption peak from MN ND to CF ND, as well as the blue shift caused by accelerated drying. Conversely, the data also indicate that slower drying promotes the formation of aggregate 1, which is likely the thermodynamically stable conformation of Y6 since it involves the largest number of bonding segments per Y6 dimer. This state is, however, less favored kinetically, as a thermal rearrangement of the molecules that leads to two overlapping segments instead of one is statistically less probable to occur. Both trends of decreasing aggregate 1 and increasing amorphous Y6 fractions, resulting in the blue shift of the Y6 absorption peak position, align with the $V_{oc}$ trend of the respective OSC devices (see Table 1), progressing from MN (lowest energy peak, lowest $V_{oc}$) to CF GQ (highest energy peak, highest $V_{oc}$). This suggests that modulating the drying speed alters the dimer configuration of Y6, and hence its optical bandgap, which consequently allows to tune the energetic properties of the final active layer.

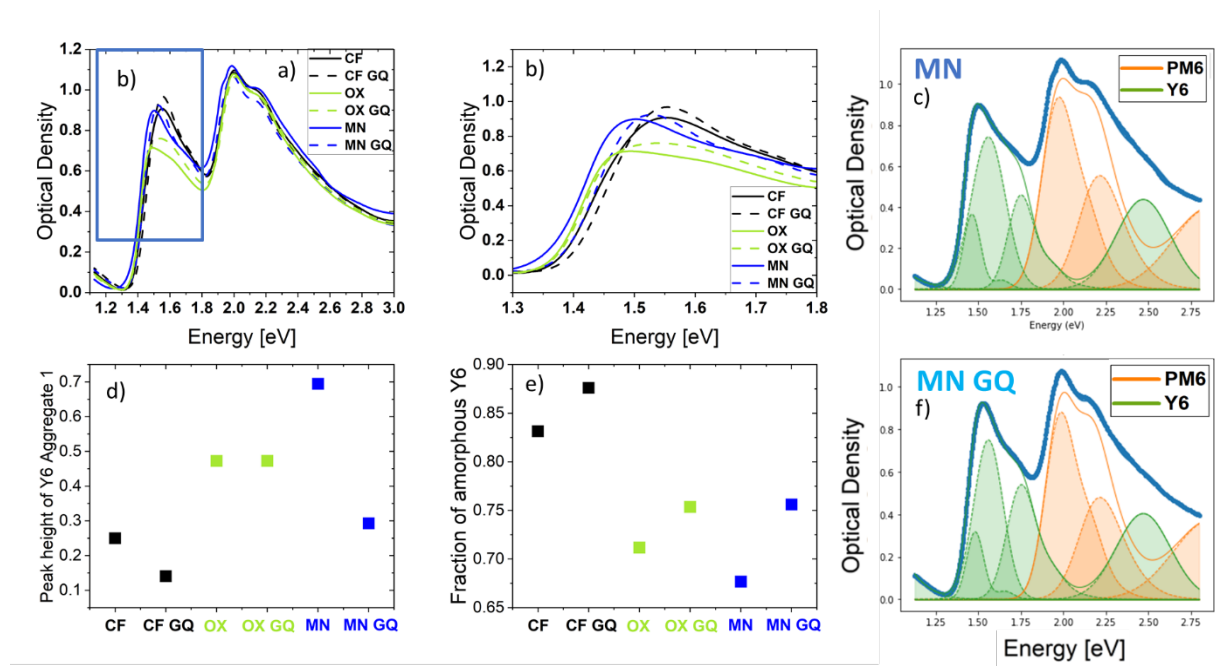

*Figure 12: a) UV-Vis spectra of different PM6:Y6 organic active layers cast from CF at 30°C (black), OX at 60°C (green) and MN at 100°C (blue), without (solid lines) and with (dotted lines) gas quenching, b) enlargement of a). Spectral modeling of the MN (c) and MN GQ spectra (f), evaluation of the simulations according to the peak height of Y6 aggregate 1 (d) and the total amorphous fraction of Y6 (e).*

## 3. Discussion

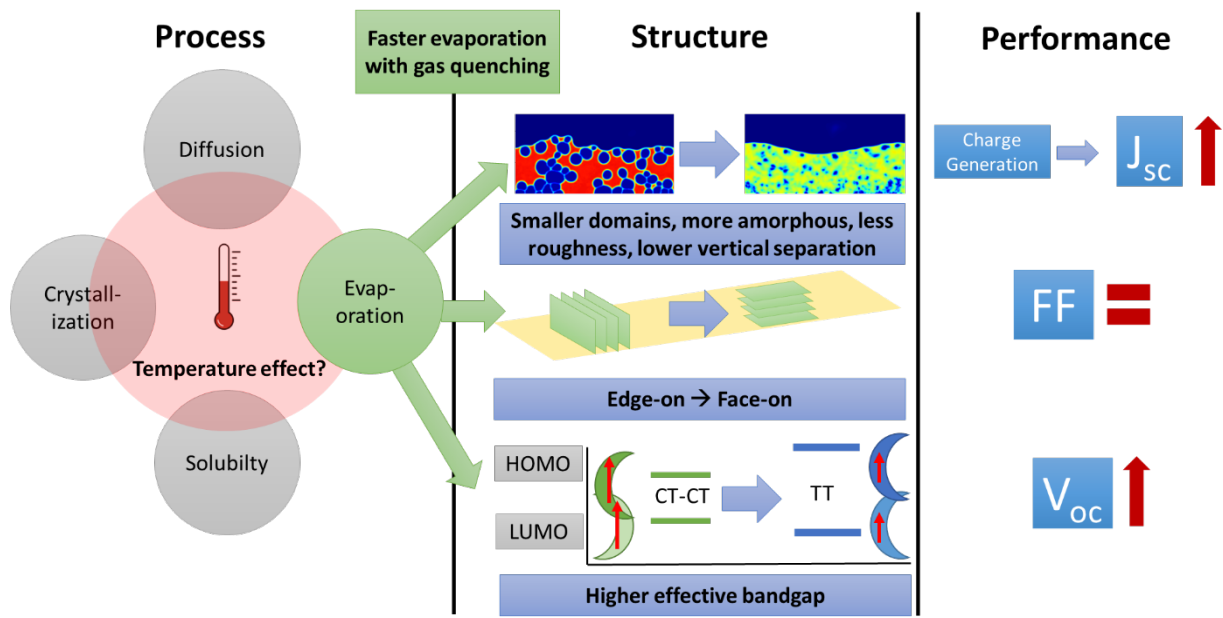

*Figure 13: Gas quenching helps to understand Process-Structure-Performance relationships for OPV bulk heterojunctions by separating the effect of the evaporation rate from the other temperature-dependent parameters. Increasing the drying rate results in a finer and more dispersed nanomorphology, which is beneficial for free charge carrier generation and, consequently, leads to improved short-circuit current densities. Additionally, for the PM6:Y6 system, the molecular stacking arrangement tends to become dominantly face-on oriented upon accelerated drying. The aggregation mode of the Y6 small molecule also changes toward a configuration associated with a larger optical bandgap, which relates to higher open-circuit voltages.*

The results demonstrate that, even without adjusting the process temperature, varying the drying rate is a simple strategy to control the organic bulk heterojunction morphology (Figure 13). As suggested by the Phase-Field simulations, fast evaporation leads to a high degree of supersaturation in the originally mixed amorphous phase, and thus to a high nucleation/growth ratio. Additionally, as the solvent is rapidly removed in parallel to the ongoing phase separation processes, the viscosity in the film quickly becomes relatively high, resulting in slower molecular diffusion kinetics. Consequently, crystalline regions remain small, emerging domains exhibit improved intermixing properties and a higher surface-to-volume ratio, and the persisting proportion of the initially mixed amorphous phase in the dry active layer is comparatively high. These features are corroborated by AFM and FLAS measurements, which show that higher drying rates give rise to reduced domain sizes, lower surface roughness, and less vertical phase segregation. The TEM mapping is also in line with the Phase-Field simulations, as accelerated evaporation induces finer spatial distribution patterns of donor and acceptor, along with smaller crystal coherence lengths overall in the dry film. This is further confirmed by the spectral decomposition performed on the UV-Vis data, where the signal pertaining to the amorphous acceptor is enhanced upon gas quenching.

The morphological changes caused by the variation of the evaporation rate reflect on the optoelectronic properties and the photovoltaic performance: Faster drying yields higher $J_{sc}$. The main reason for the improved short-circuit current density is the increased interface area between donor and acceptor, which results in an enhanced charge generation rate, as seen with the Drift-Diffusion analysis. It can be remarked that the $J_{sc}$ boost with accelerated evaporation ends in a plateau because drying rate increments no longer affect the dry active layer morphology when the kinetics of evaporation significantly exceed those of phase separation. Therefore, there is a higher relative performance gain in applying gas quenching with slower-evaporating solvents like OX and MN, as compared to CF. Note also that, depending on the properties of the material system, the intrinsic charge carrier lifetimes and the distances towards the charge transport layers, an excessive drying rate may produce too intermixed configurations. This can still increase charge generation but hinder efficient charge extraction at the same time, which may finally reduce the $J_{sc}$.

In contrast to $J_{sc}$, $V_{oc}$ variations upon gas quenching are here mainly ascribed to changes in the preferential dimer aggregation conformation that the Y6 small molecule adopts. The UV-Vis data indeed show that an increased evaporation rate tends to shift the predominant aggregation behavior from the thermodynamically-favored aggregate 1 with two bonding molecular segments towards the kinetically-favored aggregate 2 with a single overlapping segment. This latter aggregation state exhibits a larger HOMO-LUMO energy level difference, as compared to the former, and its presence correlates directly with effective bandgap and $V_{oc}$ improvements, as highlighted by the comparison between the spectral deconvolutions and the Drift-Diffusion results. The TEM investigations additionally reveal that the aggregate 2 is preferentially oriented face-on, whereas the orientation of aggregate 1 is more random. This suggests better charge transport and charge extraction properties for aggregate 2.

In the OX case, it is observed that the fraction of aggregate 1 (and thus the $V_{oc}$) does not vary significantly upon gas quenching, likely because the Y6 has undergone type 1 preaggregation already in the solution. This is supported by the FLAS results, which evidence a macroscopic vertical phase segregation with a high Y6 content at the top of the film for OX-cast samples even under fast evaporation conditions. The Phase-Field simulations attribute such a phase segregation to the presence of Y6 crystals within the wet film already at early drying stages, allowing them to progressively accumulate at the moving liquid-vapor interface until all solvent is removed.

The effect of an increased evaporation rate on the FF is less clear than the one on the $J_{sc}$ and the $V_{oc}$. Notable discrepancies arise depending on the employed solvent. For CF and OX, the total mobility and the FF seem to stagnate, or even decrease, under higher evaporation rates. This can be expected since gas quenching leaves larger amounts of amorphous regions in the bulk heterojunction, which are detrimental for charge mobility and which are also anticipated to be responsible for higher charge recombination rates. For MN, however, it is measured that the FF increases steadily with faster evaporation, and that this is mirrored by the overall charge mobility. From the perspective of the morphology, the most noteworthy MN-specific feature that can be conjectured to relate to this observation is the fact that naturally dried MN-cast active layers exhibit large leaf-shaped PM6 domains, which are suppressed upon gas quenching. Nevertheless, the reason why the presence of these PM6 domains correlates with the drastic reduction in total mobility and FF remains to be elucidated.

## 4. Summary and Conclusions

To summarize, this study aims to shed light on the role of the solvent evaporation rate in the nanomorphology formation of solution-deposited organic active layers. In comparison to previous works, where the drying kinetics were adjusted indirectly by increasing the processing temperature, they are here modulated by means of an air knife that is mounted on a blade coating device. The present approach allows to investigate drying rate variations while avoiding the presence of additional interfering phenomena that otherwise may be witnessed due to the temperature-dependence of many system properties.

This complexity is thus disentangled in this work, showing that, even at constant temperature, morphology and, consequently, optoelectronic properties and solar cell performance can be controlled to a large extent by varying the drying rate. Hence, the findings corroborate that evaporation regulation is a key lever for active layer morphology optimization. It is observed that increasing the evaporation rate generally boosts device performance, provided that the natural drying kinetics of the solvent are not already significantly faster than those of the morphology structuring mechanisms (e.g., crystallization or phase separation). For example, the supersaturation level at which crystallization phase transitions occur in the wet film can be adapted by adjusting the drying speed, and, hence, the resulting balance between crystal nucleation and growth can be modulated. This permits to generate donor and acceptor domains with adequate size, shape, and spatial distribution for improved charge separation and transport, thus increasing the short-circuit current density. Furthermore, the bulk heterojunctions produced under accelerated evaporation generally exhibit increased amounts of amorphous and intermixed material. For the Y6 small molecule, it additionally results in a change of the dominant aggregation behavior (i.e., from type 1 to type 2 aggregate form) which, in agreement with previous reports, is demonstrated to have a beneficial impact on device performance, as it leads to higher open-circuit voltages.

To finish, the overarching goal of this research is to facilitate future upscaling and industrialization of organic solar cells. With this perspective, the technique employed to tune the gas flow is designed to be readily implementable on large-scale factory equipment. As this work highlights, precise drying rate adjustment is indispensable to control morphology-defining aspects like the aggregation type, the nucleation/growth balance, and the phase separation strength during active layer fabrication. The developed gas quenching method thus constitutes an important addition to the existing processing toolkit for deriving physically-motivated design strategies for scalable organic electronics.

## 5. Acknowledgements

We gladly acknowledge Gitti L. Frey and her research group for fruitful contributions to the scientific discussion.

We acknowledge the DFG project 'Process-Structure Relationships for Solution-Processed Organic Photovoltaics' (GZ: HA4382/10-1 and BR4031/20-1) for financial support, along with the Collaborative Research Center ChemPrint (CRC1719, Project No. 538767711). We also acknowledge funding from the European Union's Horizon 2020 INFRAIA program under Grant Agreement No. 101008701 ('EMERGE') as well as the Horizon 2020 Research and Innovation Program (H2020 Societal Challenges) under grant number 952911 ('BOOSTER'). The Solar Factory of the Future (SFF) as part of the Energy Campus Nürnberg (EnCN) is acknowledged, which is supported by the Bavarian State Government (FKZ 20.2-3410.5-4-5). Part of this work has also been supported by the Helmholtz Association in the framework of the innovation platform "Solar TAP". J. Graetz and T. Unruh acknowledge the funding of the consortium DAPHNE4NFDI in the context of the work of the NDFI e.V. The consortium is funded by the DFG project No. 460248799.

## 6. Author Contributions

First authorship of this article is shared by M. Steinberger and M. Siber, as both contributed equally to the design of the investigations, to the production of results, to the scientific discussion, and to the writing of the final report. M. Steinberger was responsible for the experimental parts of this research project, while M. Siber was responsible for the theoretical modeling and simulation work.

M. Steinberger: Investigation (Photoactive layers and solar cells: fabrication and characterization), Methodology (Experimental: ND and GQ active layer fabrication, JV and TPV/CE/Photo-CELIV measurements, WLR experiments, AFM characterizations, UV-Vis acquisitions and spectral deconvolutions), Data curation, Conceptualization, Visualization, Writing – original draft, Writing – review & editing

M. Siber: Investigation (Photoactive layers and solar cells: simulation and analysis), Methodology (Theoretical: Phase-Field simulations, Drift-Diffusion fits, Quantitative evaporation rate estimation), Formal analysis (Phase-Field modeling, Drift-Diffusion modeling, Hertz-Knudsen modeling), Data curation, Conceptualization, Visualization, Writing – original draft, Writing – review & editing

H.-J. Egelhaaf: Conceptualization, Methodology, Supervision, Project administration, Funding acquisition, Resources, Writing – review & editing

M. Wu: Investigation (Active layer TEM nanomorphology characterization), Methodology (EF-SAED, STEM-EELS), Data curation, Visualization, Writing – review & editing

I. Kraus: Investigation (Active layer TEM nanomorphology characterization), Methodology (EF-SAED, STEM-EELS), Data curation, Visualization, Writing – review & editing

J. Will: Investigation (Active layer TEM nanomorphology characterization), Methodology (EF-SAED, STEM-EELS), Data curation, Visualization, Writing – review & editing

X. Xie: Investigation (FLAS), Methodology (FLAS), Writing – review & editing

L. Bu: Methodology (FLAS), Supervision, Data Curation, Writing – review & editing

J. Graetz: Methodology, Writing – review & editing

T. Unruh: Methodology, Resources, Writing – review & editing

L. Lüer: Software (Spectral deconvolution tool), Supervision, Methodology, Writing – review & editing

E. Spieker: Methodology, Supervision, Resources, Writing – review & editing

A. Distler: Conceptualization, Methodology, Supervision, Project administration, Writing – review & editing

J. Harting: Supervision, Project administration, Funding acquisition, Resources, Writing – review & editing

C.J. Brabec: Supervision, Project administration, Funding acquisition, Resources, Writing – review & editing

O.J.J. Ronsin: Conceptualization, Software (Phase-Field, Drift-Diffusion), Methodology, Supervision, Project administration, Funding acquisition, Writing – review & editing

## 7. Data Availability

In compliance with the regulations for projects funded by the German Research Foundation (DFG), the simulation data used for this article is made publicly accessible (see DOI https://doi.org/10.5281/zenodo.17951559)

## 8. Conflicts

There are no conflicts to declare.